\documentclass[preprint,prd,aps,floats,nofootinbib,showpacs,floatfix]{revtex4-1}

\usepackage{graphicx}
\usepackage{epstopdf}

\newcommand{\PRLett}[3]{\emph{Physical Review Letters}, vol. #1, Article ID #2, #3.}
\newcommand{\PRLettwithoutdot}[3]{\emph{Physical Review Letters}, vol. #1, Article ID #2, #3}
\newcommand{\PRD}[3]{\emph{Physical Review D}, vol. #1, Article ID #2, #3.}
\newcommand{\PRDwithoutdot}[3]{\emph{Physical Review D}, vol. #1, Article ID #2, #3}
\newcommand{\PRC}[3]{\emph{Physical Review C}, vol. #1, Article ID #2, #3.}
\newcommand{\PRCwithoutdot}[3]{\emph{Physical Review C}, vol. #1, Article ID #2, #3}
\newcommand{\PLB}[3]{\emph{Physics Letters B}, vol. #1, p. #2, #3.}
\newcommand{\EPJA}[3]{\emph{The European Physical Journal A}, vol. #1, p. #2, #3.}

\newcommand{\NPA}[3]{\emph{Nuclear Physics A}, vol. #1, p. #2, #3.}

\newcommand{\NPB}[3]{\emph{Nuclear Physics B}, vol. #1, p. #2, #3.}
\newcommand{\NPBwithoutdot}[3]{\emph{Nuclear Physics B}, vol. #1, p. #2, #3}

\newcommand{\RevMP}[3]{\emph{Reviews of Modern Physics}, vol. #1, p. #2, #3.}
\newcommand{\PPNP}[3]{\emph{Progress in Particle and Nuclear Physics}, vol. #1, p. #2,  #3.}

\newcommand{\gam }{\ensuremath{\gamma }}
\newcommand{\gs }{\ensuremath{\sigma }}     
\newcommand{\gz }{\ensuremath{\zeta }}       
\newcommand{\gd }{\ensuremath{\delta }}      
\newcommand{\gc }{\ensuremath{\chi }}          
\newcommand{\go }{\ensuremath{\omega }}   
\newcommand{\gr }{\ensuremath{\rho }}         
\newcommand{\gl }{\ensuremath{\lambda}}     

\newcommand{\gn }{\ensuremath{\eta}}         
\newcommand{\gp }{\ensuremath{\phi }}         

\newcommand{\gX }{\ensuremath{\Xi }}          
\newcommand{\gS }{\ensuremath{\Sigma }}   
\newcommand{\gL }{\ensuremath{\Lambda }} 
\newcommand{\gG }{\ensuremath{\Gamma }}   

\newcommand{\fd }{\ensuremath{f_{D}}}
\newcommand{\fds }{\ensuremath{f_{D_S}}}

\newcommand{\llags}[1]{\mathcal{L}_{#1}} 

\newcommand{\dmucon}{\ensuremath{\partial^\mu}} 
\newcommand{\dmucov}{\ensuremath{\partial_\mu}} 

\newcommand{\dsplus }{\ensuremath{D_S^+}}
\newcommand{\dsminus }{\ensuremath{D_S^-}}
\newcommand{\dsm }{\ensuremath{D_S} meson}
\newcommand{\dm }{\ensuremath{D} meson}

\newcommand{\mds }{\ensuremath{m_{D_S}}}

\newcommand{\lag }{Lagrangian}
\newcommand{\lagd }{Lagrangian density}

\newcommand{\wtt }{Weinberg-Tomozawa term}

\newcommand{\grz }{\ensuremath{\rho_0}}
\newcommand{\gri }{\ensuremath{\rho_i}}
\newcommand{\grsi }{\ensuremath{\rho_i^s}}
\newcommand{\grb }{\ensuremath{\rho_B}}

\newcommand{\grsub}[1]{\ensuremath{\rho_{#1}}}
\newcommand{\grssub}[1]{\ensuremath{\rho_{#1}^s}}

\newcommand{\isoap }{isospin asymmetry parameter}
\newcommand{\dg}{\dagger}
\newcommand{\chigrp}{\ensuremath{SU(3)_L\times SU(3)_R}} 

\begin{document}

%
\title[$D_S$ Mesons in Asymmetric Hot and Dense Hadronic 
Matter]{\boldmath $D_S$ Mesons in Asymmetric Hot and Dense Hadronic 
Matter}



\author{Divakar Pathak}
\email{dpdlin@gmail.com}
\affiliation{Department of Physics, Indian Institute of Technology $-$ Delhi, Hauz Khas, New Delhi $-$ 110 016, India}
\author{Amruta Mishra}
\email{amruta@physics.iitd.ac.in}
\affiliation{Department of Physics, Indian Institute of Technology $-$ Delhi, Hauz Khas, New Delhi $-$ 110 016, India}
%
\begin{abstract}
\ \\
The in-medium properties of $D_S$ mesons are investigated within 
the framework of an effective hadronic model, which is a generalization
of a chiral SU(3) model, to SU(4), in order to study the interactions
of the charmed hadrons. In the present work, the $D_s$ mesons are 
observed to experience net attractive interactions in a dense hadronic 
medium, hence reducing 
the masses of the $D_S^+$ and $D_S^-$ mesons from the vacuum values. 
While this conclusion holds in both nuclear and hyperonic media, 
the magnitude of the mass drop is observed to intensify with the 
inclusion of strangeness in the medium. Additionally, in hyperonic 
medium, the mass degeneracy of the $D_S$ mesons is observed to be 
broken, due to opposite signs of 
the Weinberg-Tomozawa interaction term in the Lagrangian density. 
Along with the magnitude of the mass drops, the mass splitting 
between $D_S^+$ and $D_S^-$ mesons is also observed to grow with 
an increase in baryonic density and strangeness content of the medium. 
However, all medium effects analyzed are found to be weakly dependent 
on isospin asymmetry and temperature. 
We discuss the possible implications emanating from this analysis, 
which are all expected to make a significant difference to observables 
in heavy ion collision experiments, 
especially the upcoming Compressed Baryonic Matter (CBM) experiment 
at the future Facility for Antiproton and Ion Research (FAIR), GSI,  
where matter at high baryonic densities is planned to be produced.  
\end{abstract}


\maketitle

\begin{section}{Introduction}

An effective description of hadronic matter is fairly common 
in low-energy QCD \cite{leshouches, Rev_1997, RMP2010}. Realizing 
that baryons and mesons constitute the effective degrees of freedom 
in this regime, it is quite sensible to treat QCD at low-energies 
as an effective theory of these quark bound states \cite{leshouches}. 
This approach has been vigorously pursued in various incarnations over the years, with the different adopted strategies representing merely different manifestations of the same underlying philosophy. The actual manifestations range from the quark-meson coupling model \cite{QMC1, QMC2}, phenomenological, relativistic mean-field theories based on the Walecka model \cite{serotwalecka}, along with their subsequent extensions, the method of QCD sum rules \cite{QSR_original_and_Shifman_review}, as well as the coupled channel approach \cite{Oset_PRL98, Oset_NPA98} for treating dynamically generated resonances, which has further evolved into more specialized forms, namely the local hidden gauge theory \cite{Hidden_local_gauge_formalism_Refs1,Hidden_local_gauge_formalism_Refs2}, as well as formalisms 
based on incorporating 
heavy-quark spin symmetry (HQSS) \cite{Heavy_Quark_Spin_Symmetry_Refs1, 
Heavy_Quark_Spin_Symmetry_Refs2} 
into the coupled channel framework 
\cite{HQSS_Example_Papers1,HQSS_Example_Papers2,HQSS_Example_Papers3,
HQSS_Example_Papers4, HQSS_Example_Papers5}. 
Additionally, the method of chiral invariant \lag s \cite{leshouches} 
(which shall also be embraced in this work), has developed over the years 
into a very successful strategy. The same constitutes an effective 
field theoretical model in which the specific form of hadronic interactions 
is dictated by symmetry principles, and the physics governed predominantly 
by the dynamics of chiral symmetry -- its spontaneous breakdown implying 
a non-vanishing scalar condensate $\langle {\bar q}q \rangle$ in  vacuum. 
One naturally expects then, that the hadrons composed of these quarks 
would also be modified in accord with these condensates 
\cite{Rev_1997, RMP2010}. But while all hadrons would be subject 
to medium modifications from this perspective, pseudoscalar mesons 
have a special role in this context. In accordance with the 
Goldstone's theorem \cite{leshouches}, spontaneous breaking 
of chiral symmetry leads to the occurrence of massless pseudoscalar 
modes, the so-called Goldstone bosons, which are generally identified 
with the spectrum of light pseudoscalar mesons, like the pions, 
or kaons and anti-kaons \cite{Rev_1997, RMP2010}. In a strict sense, 
however, none of these physical mesons is a true Goldstone mode, 
since they are all massive, while Goldstone modes are supposed 
to be massless \cite{Rischke_QGP_Review}. The origin of the masses 
of these mesons is related to the non-zero masses of light quarks, 
as can be easily discerned from the Gell-Mann-Oakes-Renner (GOR) 
relations, and hence, from explicit symmetry breaking terms 
(or explicit mass terms) in the chiral effective framework 
\cite{leshouches}. In fact, if one considers the limiting situation 
of vanishing quark masses $\left( m_q \rightarrow 0 \right)$, 
the masses of these pseudoscalar mesons would also vanish, 
so that the perfect Goldstone modes are indeed recovered. 
For this reason, the physically observed light pseuodscalar 
mesons are dubbed pseudo-Goldstone bosons \cite{Rischke_QGP_Review}. 
In purely this sense, therefore, there is an inherent similarity 
between all these classes of pseudoscalar mesons -- masses are 
acquired through explicit quark mass 
terms, and only the magnitude of these mass terms differ 
between these cases, being small for the pions (since $m_u, m_d < 10$ MeV), 
comparatively larger for the kaons and antikaons (since $m_s \sim 150$ MeV),
and appreciably larger for the charmed pseudoscalar mesons. Thus, 
as we advance from the pions to the strange pseudoscalar mesons, 
with increasing mass, these mesons depart more from the ideal 
Goldstone mode character pertaining to the theorem and 
there is considerable departure of the charmed pesudoscalar 
mesons from Goldstone mode behaviour due to the explicit chiral symmetry
breaking arising from the large charm quark mass ($m_c \simeq$ 1.3 GeV).
An understanding of the in-medium properties 
of the pseudoscalar mesons has been an important topic of research, 
both theoretically and experimentally. 
Within the chiral effective approach, the pseudoscalar mesons
are modified in the medium due to the modifications of the
quark condensates in the hadronic medium.
For pions, it is observed however, 
that medium effects for them are weakened by the smallness of 
explicit symmetry breaking terms \cite{Rev_1997, RMP2010}. 
Considerably detailed analysis of medium effects have been performed 
over the years, particularly in a chiral SU(3) approach, 
for strange pesudoscalar mesons (kaons and antikaons) 
\cite{sambuddha1, sambuddha2, mam_kaons2006, mam_kaons2008}. 
For studying the charmed mesons, one needs to generalize the
SU(3) model to SU(4), in order to incorporate the interactions 
of the charmed mesons to the light hadrons.
Such a generalization from SU(3) to SU(4) was initially
done in Ref. \cite{gamermann_oset}, where the interaction 
Lagrangian was constructed for the pesudoscalar mesons
for SU(4) from a generalization of the lowest order chiral SU(3) 
Lagrangian. Since the chiral symmetry is explicitly broken for 
the SU(4) case due to the large mass of the charm quark 
($m_c \simeq$1.3 GeV),
which is much larger than the masses of the light quarks,
for the study of the charmed ($D$) pseudoscalar mesons
\cite{mamD2004,arindam,arvDprc}, we adopt 
the philosophy of generalizing the chiral SU(3) model to SU(4) 
to derive the interactions of these mesons with the light hadrons,
but use the observed masses of these heavy hadrons as well as
empirical/observed values of their decay constants 
\cite{liukolin}.  
With all these studies proving to be informative, 
the most natural direction of extension of  this approach 
would be to analyze these medium effects for a \emph{strange-charmed} 
system (the \dsm s). Apart from pure theoretical interest, 
an understanding of the medium modifications of \dsm s is important, 
since these can make a considerable difference to experimental 
observables in the (ongoing and future) relativistic heavy ion 
collision experiments, 
besides being significant in questions concerning their production and transport in such experimental situations. For instance, in a recent work, He et al. \cite{friesDs} have shown that the modifications of the \dsm \ spectrum can serve as a useful probe for understanding key issues regarding hadronization in heavy ion collisions. It is suggested that by comparing observables for $D$ and \dsm s, it is possible to constrain the hadronic transport coefficient. This comparison is useful since it allows a clear distinction between hadronic and quark-gluon plasma behavior. 

However, as far as the existing literature on this strange-charmed system of mesons is concerned, we observe that only the excited states of \dsm s have received considerable attention, predominantly as dynamically generated resonances in various coupled channel frameworks \cite{excited_ds_PPNP_review,excited_ds_review}. One must bear in mind that in certain situations, a molecular interpretation of these excited states (resonances), is more appropriate \cite{close_nature_2003} for an explanation of their observed, larger than expected lifetimes. From this perspective, a whole plethora of possibilities have been entertained for the excited $D_S$ states, the standard quark-antiquark picture aside. These include their description as molecular states borne out of two mesons, four-quark states, or the still further exotic possibilities - as two-diquark states and as a mixture of quark-antiquark and tetra-quark states \cite{excited_ds_PPNP_review}. Prominent among these is a treatment of $D^*_{S0}(2317)$ as a $DK$ bound state \cite{simonov,wang_and_wang}, that of $D_{s1}(2460)$ as a dynamically generated $D^* K$ resonance \cite{ds2460_plb2007}, $D^*_{S2}(2573)$ being treated within the hidden local gauge formalism in coupled channel unitary approach \cite{oset_ds2573,Oset_conf_ds_molecular}, as well as the vector $D_S^*$ states and the $D_S^+(2632)$ resonance treated in a multichannel approach \cite{prl_ds2632_2004}. The former three have also been covered consistently under the four-quark picture \cite{ds_4quark_prl2004}. So, while considerable attention has been paid to dynamically generating the higher excited states of the \dsm s, there is a conspicuous dearth \cite{lutzkorpa} of available information about the medium modifications of the lightest pseudoscalar \dsm s, ${D_S}(1968.5)$, the one that we know surely, is well described within the 
quark-antiquark picture. 
In fact, to the best of our knowledge, the entire existing literature about the $(J_P = 0^-)$ \dsm s in a hadronic medium is limited to the assessment of their spectral distributions and medium effects on the dynamically generated resonances borne out of the interaction of these \dsm s with other hadron species, in the coupled channel analyses of Refs. \cite{lutzkorpa, hofmannlutz, JimenezTejero_2009vq, JimenezTejero_2011fc}. Compared with their open-charm, non-strange counterparts, the $D$ mesons, which have been extensively investigated using a multitude of approaches over the years and consequently boast of an extensive literature, the literature concerning the medium behavior of \dsm s can at best be described as scanty, and there is need for more work on this subject.    
If one considers this problem from the point of view of the (extended) chiral effective approach, this 
scantiness is most of all, because of the lack of a proper framework where the relevant form of the interactions for the \dsm s with the light hadrons (or in more generic terms, of meson-baryon interactions with the charm sector covered), based on arguments of symmetry and invariance, could be written down. Clearly, such interactions would have to be based on $SU(4)$ symmetry and bear all these pseudoscalar mesons and baryons in 15-plet and 20-plet representations respectively, with meson-baryon interaction terms  
still in accordance with the general framework for writing chiral-invariant structures, as well as bearing appropriate symmetry breaking terms obeying the requisite transformation behavior under chiral transformations \cite{leshouches}, which is quite a non-trivial problem.  
Of late, such formalisms have been proposed in Refs. \cite{hofmannlutz,arvDepja}, as an extension of the frameworks based on chiral \lag s, where $SU(4)$ symmetry 
forms the basis for writing down the relevant interaction terms. However, 
since the mass of the charm quark is approximately $1.3$ GeV \cite{PDG2012}, 
which is considerably larger than that of the up, down and strange quarks, 
the $SU(4)$ symmetry is explicitly broken by this large charm quark mass. 
Hence, this formalism only uses the symmetry to derive the form of the interactions, whereas an explicit symmetry breaking term accounts for the large quark mass through the introduction of mass terms of the relevant ($D$ or $D_S$) mesons. Also, $SU(4)$ symmetry being badly broken implies that any symmetry and order in the masses and decay constants, as predicted on the basis of $SU(4)$ symmetry, would not hold in reality. The same is acknowledged in this approach 
\cite{arvDepja} and as has been already mentioned, one does not use 
the masses and decay constants as expected on the basis of $SU(4)$ symmetry, 
but rather, their observed, Particle Data Group (PDG) \cite{PDG2012} values. 
Overall, therefore, $SU(4)$ symmetry is treated (appropriately) as being 
broken in this approach. 
Also, it is quite well known, both through model-independent \cite{roder} and model-dependent \cite{arvDprc} calculations that the light quark condensates $\left( \langle {\bar u} u \rangle, \langle {\bar d} d \rangle \right)$ are modified significantly in a hadronic medium with medium parameters like density and temperature, the strange quark condensate $\langle {\bar s} s \rangle$ is comparatively stolid and its variation is significantly more subdued, while upon advancing to the charm sector, the variation in the charmed quark condensate $\langle {\bar c} c \rangle$ is altogether negligible in the entire hadronic phase \cite{roder}. 
These observations form the basis for treating the charm degrees of freedom of open charm pseudoscalar mesons as frozen in the medium, as was the case in the treatments of Refs. \cite{arindam, arvDprc, arvDepja}. Thus, as we advance from pions and kaons to the charmed pseudoscalar mesons, the generalization is perfectly natural but with the aforementioned caveats. 
Provided all these aspects are taken into account, a generalization of this chiral effective framework to open charm pseudoscalar mesons is quite reasonable and sane, and the predictions from such an extended chiral effective approach bodes very well \cite{arindam} with alternative calculations based on the QCD sum rule approach, quark meson coupling model, coupled channel approach, as well as studies of quarkonium dissociation using heavy-quark potentials from lattice QCD at finite temperatures. Additionally, it is interesting to note that this approach, followed in Refs. \cite{arindam, arvDprc, arvDepja} for the charmed pseudoscalar mesons, has recently been extended to the bottom sector and used to study the medium behavior of the open bottom pseudoscalar $B$, ${\bar B}$ and $B_S$ mesons \cite{DP_Bm_PRC2015, DP_Bsm_IJMPE}. The inherent philosophy beneath this extension continues to be the same -- the dynamics of the heavy quark/anti-quark is treated as frozen, and the interactions of the light quark (or anti-quark) of the meson, with the particles constituting the medium, are responsible for the medium modifications. With this subsequent generalization as well, the physics of the medium behavior that follows from this approach, is in agreement \cite{DP_Bm_PRC2015} with works based on alternative, independent approaches, like the heavy meson effective theory, quark-meson coupling model, as well as the QCD sum rule approach. Thus, these aforementioned, prior works based on the generalization of the original chiral effective approach to include heavy flavored mesons, are totally concordant with results from alternative approaches followed in the literature, which lends an aura of credibility to this strategy. 
Given this backdrop, it is clear that  
these formalisms wipe out the reason why such an investigation 
for the \dsm s within the effective hadronic model,
obtained by generalizing the chiral SU(3) model to SU(4),
has not been undertaken till date, and permit this attempt 
to fill the void.

We organize this article as follows: in section II, we outline the Chiral \chigrp \ Model (and its generalization to the $SU(4)$ case) used in this investigation. In section III, the \lagd \ for the \dsm s, within this extended framework, is explicitly written down, and is used to derive their in-medium dispersion relations. In section IV, we describe and discuss our results for the in-medium properties of \dsm s, first in the nuclear matter case, and then in the hyperonic matter situation, following which, we briefly discuss the possible implications of these medium modifications. 
Finally, we summarize the entire investigation in section 
V.
\label{section1}
\end{section}

\begin{section}{The Effective Hadronic Model}
As mentioned previously, this study is based on a generalization of the chiral \chigrp \ model \cite{Pap_prc99}, to $SU(4)$. We summarize briefly the rudiments of the model, while referring the reader to Refs. \cite{Pap_prc99,Zsch} for the details. This is an effective hadronic model of interacting baryons and mesons, based on a non-linear realization of chiral symmetry \cite{weinberg67,weinberg68}, where chiral invariance is used as a guiding  principle, in deciding the form of the interactions \cite{coleman1,coleman2,bardeenlee}. 
Additionally,  the model incorporates a scalar dilaton field, $\chi$, to mimic the broken scale invariance of QCD \cite{Zsch}. Once these invariance arguments determine the form of the interaction terms, one resorts to a phenomenological fitting of the free parameters of the model, to arrive at the desired effective Lagrangian density for these hadron-hadron interactions.  The general expression for the chiral model \lagd \ reads: 
\begin{equation}
{\cal L} = {\cal L}_{\rm kin}+{\sum_{W}} {\cal L}_{\rm BW} + {\cal L}_{\rm vec} 
+ {\cal L}_{0} + {\cal L}_{\rm scale \; break}+ {\cal L}_{\rm SB}
\label{genlag_model}
\end{equation}
In eqn.(\ref{genlag_model}), $\llags{\rm kin}$ is the kinetic energy term, while $\llags{\rm BW}$ denotes the baryon-meson interaction term. Here, baryon-pseudoscalar meson interactions generate the baryon masses. $\llags{\rm vec}$ treats the dynamical mass generation of the vector mesons through couplings with scalar mesons. The self-interaction terms of these mesons are also included in this term. $\llags{\rm 0}$ contains the meson-meson interaction terms, which induce spontaneous breaking of chiral symmetry. $\llags{\rm scalebreak}$ introduces scale invariance breaking, via a logarithmic potential 
term in the scalar dilaton field, $\chi$. Finally, $\llags{\rm SB}$ refers to the explicit symmetry breaking term. This approach has been employed extensively to study the in-medium properties of hadrons, particularly pseudoscalar mesons  \cite{mam_kaons2006,mam_kaons2008,sambuddha1,sambuddha2}. 
As was observed in section I as well, this would be most naturally extended to the charmed (non-strange and strange) pseudoscalar mesons. However, that calls for this chiral $SU(3)$ formalism to be generalized to $SU(4)$, which has been addressed in \cite{arvDepja, hofmannlutz}. For studying the in-medium behavior of pseudoscalar mesons, the following contributions need to be analyzed \cite{arvDepja, arindam, arvDprc}: 
\begin{equation}
\llags{\ } = \llags{\rm WT} + \llags{\rm 1^{st} Range} + \llags{\rm d_1} + \llags{\rm d_2} + \llags{\rm SME}
\label{genlag}
\end{equation}
In eqn.(\ref{genlag}), $\llags{\rm WT}$ denotes the \wtt, given by the expression \cite{arvDepja} - 
\begin{eqnarray}
\llags{\rm WT} & = & -\frac{1}{2}\big[ \bar{B}_{ijk} \,\gam^\mu\,\left( ({\gG_{\mu}})_l^{\ k}\,B^{ijl}+ 2\ ({\gG_{\mu}})_l^{\ j}\,B^{ilk}\right) \big],
\label{LWTexpression}
\end{eqnarray}
with repeated indices summed over. Baryons are represented by the tensor $B^{ijk}$, which is antisymmetric in its first two indices 
\cite{hofmannlutz}. 
The indices $i, \ j$ and $k$ run from 1 to 4, and one can 
directly read the quark content of a baryon state, with the identification: 
$1 \leftrightarrow u, 2 \leftrightarrow d, 
3 \leftrightarrow s, 4 \leftrightarrow c$. However, the heavier, charmed baryons are discounted from this analysis. In eqn.(\ref{LWTexpression}), $\gG_\mu$ is defined as - 
\begin{equation}
\gG_\mu \ = -\frac{i}{4}\Big( u^{\dg} \big( \dmucov u \big) - \big( \dmucov u^{\dg} \big) u + u \big( \dmucov u^{\dg} \big) - \big( \dmucov u \big) u^{\dg} \Big),
\label{Gammamu1}
\end{equation}
where the unitary transformation operator, 
$u = \exp (iM \gamma_5/\sqrt{2} \gs_0$), is defined in terms of 
the matrix of pseudoscalar mesons, 
$M = (M^a \gl_a / \sqrt{2})$, $\gl_a$ representing the generalized 
Gell-Mann matrices. Further, $\llags{\rm SME}$ is the scalar meson 
exchange term, which is obtained from the explicit symmetry breaking term - 
\begin{equation}
\llags{\rm SB}  =  -\frac{1}{2} \ {\rm Tr} \left( A_p \left(uXu + u^{\dg}Xu^{\dg}\right) \right),
\label{LSBexpression}
\end{equation}
where $A_p = (1/\sqrt{2}) \ {\rm diag} \left[ m_{\pi}^2 f_{\pi}, m_{\pi}^2 f_\pi, \left(2 m_K^2 f_K -m_{\pi}^2 f_\pi \right), \left(2 m_D^2 f_D - m_{\pi}^2 f_\pi \right) \right]
$, and $X$ refers to the scalar meson multiplet \cite{arindam}. 
Also, the first range term is obtained from the kinetic energy term of the pseudoscalar mesons, and is given by the expression:
\begin{equation}
\llags{\rm 1^{\rm st} Range} =  {\rm Tr} \big(u_{\mu} X u^{\mu}X +X u_{\mu} u^{\mu} X \big) 
\label{Lfirstrangeexpression}
\end{equation}
where $u_{\mu} =-i \left( \big( u^{\dg} \big( \dmucov u \big)
 - \big( \dmucov u^{\dg} \big) u\big ) 
- \big ( u \big( \dmucov u^{\dg} \big)
 - \big( \dmucov u \big) u^{\dg} \big) \right)/4$.
 Lastly, the $d_1$ and $d_2$ range terms, are: 
\begin{equation}
\llags{\rm d_1} = \frac{d_{1}}{4}\Big( \bar B_{ijk} B^{ijk}{(u_\mu)}_{l}^{\ m}{(u^\mu)}_{m}^{\ l}\Big)
\label{Ld1expression}
\end{equation}
\begin{equation}
\llags{\rm d_2} =\frac{d_{2}}{2}\Big[ \bar B_{ijk} {(u_\mu)}_{l}^{\ m} \ 
\Big({(u^\mu)}_{m}^{\ k}B^{ijl} + 2{(u^\mu)}_{m}^{\ j}B^{ilk}\Big)\Big] 
\label{Ld2expression}
\end{equation}
Adopting the mean field approximation \cite{serotwalecka,Zsch}, the effective \lagd \ for scalar and vector mesons simplifies; the same is used subsequently, to derive the equations of motion for the non-strange scalar-isoscalar meson \gs, scalar-isovector meson \gd, strange scalar meson \gz, as well as for the vector-isovector meson \gr, non-strange vector meson \go \ and the strange vector meson $\phi$, within this model. 

The \dsm \ interaction \lagd \ and in-medium dispersion relations, as they follow from the above general formulation, are described next. 
\end{section}

\begin{section}{\boldmath $D_S$ Mesons in Hadronic Matter}
The \lagd \ for the \dsm s in isospin-asymmetric, strange, hadronic medium is given as -
\begin{equation}
\llags{\rm total} = \llags{\rm free} + \llags{\rm int} 
\end{equation}
This $\llags{\rm free}$ is the free \lagd \ for a complex scalar field (which corresponds to the \dsm s in this case), and reads: 
\begin{equation}
\llags{\rm free} = \left( \dmucon \dsplus\right) \left(\dmucov \dsminus\right) \ -\mds^2 \left( \dsplus \dsminus \right) 
\end{equation}
On the other hand, $\llags{\rm int}$ is determined to be:
\begin{eqnarray}
\llags{\rm int} & = & -\frac {i}{4\fds^2}\big[\Big( 2\left( \bar \gX^0 \gam^{\mu} \gX^0 + \bar \gX^- \gam^{\mu}\gX^- \right) + \bar{\gL}^{0}\gam^{\mu}\gL^{0} + \bar{\gS}^{+}\gam^{\mu}\gS^{+} \nonumber\\ 
& & \ \ \ \ \ \ \ + \ \bar{\gS}^{0}\gam^{\mu}\gS^{0} + \bar{\gS}^- \gam^{\mu} 
\gS^- \Big) \Big(\dsplus (\dmucov \dsminus) - (\dmucov \dsplus) \dsminus \Big) 
\big] \nonumber\\ 
& & + \frac{\mds^2}{\sqrt2 \fds} \big[(\gz' +\gz'_c) \left( \dsplus \dsminus \right) \big] \nonumber\\ 
& & - \frac {\sqrt2}{\fds}\big[(\gz' +\gz'_c)\ \Big( (\dmucov \dsplus)(\dmucon \dsminus)\Big)\big] \nonumber \\
& & + \frac {d_1}{2 \fds^2}\big[\Big( \bar p p +\bar n n +\bar{\gL}^{0}\gL^{0}+\bar{\gS}^{+}\gS^{+}+\bar{\gS}^{0}\gS^{0}\nonumber\\
& & \ \ \ \ \ \ \ \ \ \ \ \ \ \ + \ \bar{\gS}^{-}\gS^{-} + \bar{\gX}^{0}\gX^{0}+\bar{\gX}^{-}\gX^{-}\Big)  \Big( (\dmucov \dsplus)(\dmucon 
\dsminus) \Big)\big]\nonumber \\
& & + \frac {d_2}{2 \fds^2} \big [\Big( 2\left( \bar \gX^0 \gX^0 + \bar \gX^- \gX^- \right) + \bar{\gL}^{0}\gL^{0}+\bar{\gS}^{+}\gS^{+} \nonumber\\ 
& & \ \ \ \ \ \ \ \ \ \ \ \ \ \ \ \ \ \ \ \ \ +\  \bar{\gS}^{0}\gS^{0} + \bar{\gS}^- \gS^-\Big)\Big((\dmucov \dsplus)(\dmucon \dsminus)\Big)\big] 
\label{L_int_ds}
\end{eqnarray}
In this expression, the first term (with coefficient $-i/4\fds^2$) is the \wtt, obtained from eqn.(\ref{LWTexpression}), the second term (with coefficient $\mds^2/\sqrt{2}\fds$) is the scalar meson exchange term, obtained from the explicit symmetry breaking term of the \lag \ (eqn.(\ref{LSBexpression})), third term (with coefficient $-\sqrt{2}/\fds$) is the first range term (eqn.(\ref{Lfirstrangeexpression})) and the fourth and fifth terms (with coefficients $(d_1/\fds^2)$ and $(d_2/\fds^2)$, respectively) are the $d_1$ and $d_2$ terms, calculated from eqns.(\ref{Ld1expression}) and (\ref{Ld2expression}), respectively. Also, $\gz' = \gz - \gz_0$, is the fluctuation of the strange scalar field from its vacuum value. 
The mean-field approximation, mentioned earlier, is a useful, simplifying measure in this context, since it permits us the following replacements: 
\begin{eqnarray}
{\bar B_i} B_j \rightarrow & \langle{\bar B_i} B_j\rangle & 
\equiv \gd_{ij} \grssub{i} \label{MFA_baryons_1}\\
{\bar B_i}\gam^{\mu}B_j \rightarrow & \langle{\bar B_i}\gam^{\mu}B_j\rangle & = \gd_{ij} \left(\gd^0_{\mu} ({\bar B_i}\gam^{\mu}B_j) \right) \equiv \gd_{ij} \grsub{i}
\label{MFA_baryons}
\end{eqnarray}
Thus, the interaction \lag \ density can be recast in terms 
of the baryonic number densities and scalar densities,
given by the following expressions:
\begin{eqnarray}
\gri & = & \frac{\gamma_s}{\left( 2 \pi \right)^3} {\int {d^3 k}}
~\Bigg( \frac{1}{\exp{\left( \frac{E^*_i(k) - \mu_i^*}{T}\right)} + 1}
- \frac{1}{\exp{\left( \frac{E^*_i(k) + \mu_i^*}{T}\right)} + 1}\Bigg ),
\label{numberdensityintegral} \\
\grsi & = & \frac{\gamma_s}{\left( 2 \pi \right)^3} 
{\int {d^3 k}~\frac{m^*_i}{E^*_i (k)}}~
~\Bigg( \frac{1}{\exp{\left( \frac{E^*_i(k) - \mu_i^*}{T}\right)} + 1}
+ \frac{1}{\exp{\left( \frac{E^*_i(k) + \mu_i^*}{T}\right)} + 1}\Bigg ).
\label{scalardensityintegral}
\end{eqnarray}
In the above, $m_i^*$ and $\mu_i^*$ are the effective mass 
and effective chemical potential of the $i^{th}$ baryon, given as,
$m^*_i = -\left( g_{\gs i}\gs + g_{\gz i}\gz + g_{\gd i}\gd \right)$,
and $\mu_i^* = \mu_i - \left( g_{\gr i}\tau_3 \gr 
+ g_{\go i}\go + g_{\gp i}\gp \right)\;\;$,
$E^*_i(k)=(k^2+{m_i^*}^2)^{1/2}$ and
$\gamma_s = 2$ is the spin degeneracy factor.  
One can find the equations of motion for the \dsplus \ and \dsminus \ mesons, by the use of Euler-Lagrange equations on this \lagd. The linearity of these equations follows from eqn.(\ref{L_int_ds}), which allows us to assume plane wave solutions $( \sim \ e^{i(\vec{k}.\vec{r} - \go t)} )$, and hence, `Fourier transform' these equations, to arrive at the in-medium dispersion relations for the \dsm s. These have the general form:
\begin{equation}
-\go^{2} + \vec{k}^2 + \mds^2 - \Pi (\go,|\vec{k}| ) = 0
\label{dispersion}
\end{equation}
where, \mds \ is the vacuum mass of the \dsm s and $\Pi (\go,| \vec{k} |)$ is the 
self-energy of the \dsm s in the medium. Explicitly, the latter reads:
\begin{eqnarray}
\Pi (\go, |\vec k|) &=& \big[ \Big( \frac{d_1}{2\fds^2} \big( \grssub{p} 
+ \grssub{n} + \grssub{\gL} + \grssub{\gS^+} + \grssub{\gS^0} + \grssub{\gS^-} + \grssub{\gX^0} + \grssub{\gX^-}\big) \Big) \nonumber\\
 & & \ \ + \Big( \frac{d_2}{2\fds^2} \big( 2(\grssub{\gX^0} + \grssub{\gX^-}) + \grssub{\gL} + \grssub{\gS^+} + \grssub{\gS^0} + \grssub{\gS^-}\big) \Big) \nonumber \\
& & \ \ -\Big( \frac{\sqrt{2}}{\fds}\big( \gz' + \gz_c' \big) \Big) \big]
\Big( \go ^2 - {\vec k}^2 \Big) \nonumber\\
& & \pm \big[ \frac{1}{2\fds^2} \Big( 2(\grsub{\gX^0} + \grsub{\gX^-}) 
+ \grsub{\gL} + \grsub{\gS^+} + \grsub{\gS^0} + \grsub{\gS^-} \Big) 
\big] \go \nonumber\\
& & + \big[ \frac{\mds^2}{\sqrt{2}\fds}\big( \gz' + \gz_c' \big) \big]
\label{selfenergy}
\end{eqnarray}
where the $+$ and $-$ signs in the coefficient of \go , refer to \dsplus \ and \dsminus \  respectively, and we have used equations (\ref{MFA_baryons_1}) and (\ref{MFA_baryons}) to simplify the bilinears. In the rest frame of these mesons (i.e. setting $\vec{k} = 0$), these dispersion relations reduce to:
\begin{equation}
-\go^2 + \mds^2 - \Pi \left( \go, 0 \right) = 0,
\label{disp_relatn_k0_quad}
\end{equation}
which is a quadratic equation in \go, i.e. of the form $A\go^2 + B\go + C = 0$, where the coefficients $A, \ B$ and $C$ depend on various interaction terms in Eq. (\ref{dispersion}), and read: 
\begin{eqnarray} 
A & = & \Big[ 1 + \Big( \frac{d_1}{2\fds^2} {\sum_{(N+H)}}\grssub{i} \Big) 
+ \Big( \frac{d_2}{2\fds^2} \Big( 2 {\sum_{\gX}}\grssub{i} \ 
+ {\sum_{(H-\gX)}}\grssub{i} \Big) \Big) \nonumber\\ 
& & \ \ \ -\Big( \frac{\sqrt{2}}{\fds}\big( \gz' + \gz_c' \big) 
\Big) \Big] 
\ \ \ 
\label{disp_a}
\end{eqnarray}
\begin{equation}
B =  \pm \Big[ \frac{1}{2\fds^2} \Big(  2 {\sum _{\gX}}\grsub{i} \ 
+ {\sum _{(H-\gX)}}\grsub{i} \Big) \Big] 
\label{disp_b}
\end{equation}
\begin{equation}
C = \Big[-\mds^2 +  \frac{\mds^2}{\sqrt{2}\fds}
\big( \gz' + \gz_c' \big) \Big]
\label{disp_c}
\end{equation}
As before, the $`+'$ and $`-'$ signs in $B$, correspond to \dsplus \ and \dsminus, respectively. In writing the above summations, we have used the following notation: $H$ denotes the set of all hyperons, $N$ is the set of nucleons, $\gX$ represents the Xi hyperons  $(\gX^{-,0})$ and $(H-\gX)$ denotes all hyperons other than Xi hyperons, i.e. the set of baryons $(\gL, \ \gS^{+,-,0})$ which all carry one strange quark. 
This form is particularly convenient for later analysis. Also, the optical potential of \dsm s, is defined as:
\begin{equation}
U(k) = \go(k) - {\sqrt {k^2 + \mds^2}}
\label{OptPot_Def}
\end{equation}
where $k \ (= |{\vec k}|)$, refers to the momentum of the respective \dsm, and $\go(k)$ represents its momentum-dependent in-medium energy. 

In the next section, we study the sensitivity of the \dsm \ effective mass 
on various characteristic parameters of hadronic matter, viz. baryonic 
density (\grb), temperature ($T$), \isoap \ $\gn \ = -{\sum _{i}}  I_{3i} 
\rho_{i}/\grb$, and the strangeness fraction 
$f_s = {\sum _{i}}  |S_{i}| \rho_{i}/\rho_{B}$, where $S_{i}$ and $I_{3i}$ denote the strangeness quantum number, and the third component of isospin of the $i^{\rm th}$ baryon, respectively. 
\end{section}

\begin{section}{Results and Discussion} \label{results_section}
Before describing the results of our analysis of \dsm s in a hadronic 
medium, we first discuss our parameter choice and the various simplifying 
approximations employed in this investigation. The parameters of the 
effective hadronic model are fitted to the vacuum masses of baryons, 
nuclear saturation properties and other vacuum characteristics 
in the mean field approximation \cite{Pap_prc99,Zsch}. 
In this investigation, we have used the same parameter set that 
has earlier been used to study charmed ($D$) mesons within 
this effective hadronic model \cite{arvDepja}. 
In particular, we use the same values of the parameters $d_1$
and $d_2$ of the range terms ($d_1=2.56/m_K$ and $d_2=0.73/m_K$),
fitted to empirical values of kaon-nucleon 
scattering lengths for the $I=0$ and $I=1$ channels, as were employed 
in earlier treatments \cite{arvDepja, arindam, sambuddha1, sambuddha2}. 
For an extension to the strange-charmed system, the only extra parameter 
that needs to be fitted is the \dsm \ decay constant, \fds, which is 
treated as follows. 
%
By extrapolating the results of Ref.\cite{bardeenlee}, we arrive at the following expression for \fds \ in terms of the vacuum values of the strange and charmed scalar fields: 
\begin{equation}
\fds \ = \ \ \ \frac{-\left( \gz_0 + {\gz_c}_0 \right)}{\sqrt{2}}.
 \label{fds_exp}
\end{equation}
For fitting the value of \fds, we retain the same values of $\gs_0$ and $\gz_0$ as in the earlier treatments of this chiral effective model \cite{Pap_prc99,sambuddha2}, and 
determine ${\gz_c}_0$ from the expression from the expression for \fd \ in terms of $\gs_0$ and ${\gz_c}_0$ \cite{arindam}, using the Particle Data Group (PDG) value of $\fd = 206$ MeV. 
Substituting these in eqn.(\ref{fds_exp}), our fitted value of \fds \ comes out to be 235 MeV, which is close to its PDG value of 260 MeV, and particularly, is of the same order as typical lattice QCD calculations for the same \cite{PDG2012}. We therefore persist with this fitted value $\fds = 235$ MeV in this investigation. 
Also, we treat the charmed scalar field ($\gz_c$) as being arrested 
at its vacuum value (${\gz_c}_0$) in this investigation, as has been 
considered in other treatments of charmed mesons in a hadronic context 
\cite{arindam,arvDepja}. This neglect of charm dynamics appears natural 
from a physical viewpoint, owing to the large mass of a charm quark 
($m_c \sim 1.3$ GeV) \cite{PDG2012}. The same was verified in an 
explicit calculation in Ref.\cite{roder}, where the charm condensate was observed to vary very weakly in the temperature range of interest to us in this regime \cite{LQCD1,LQCD2,LQCD3}. As our last approximation, we point out that in the current investigation, we work within the \emph{`frozen glueball limit'} \cite{Zsch}, where the scalar dilaton field (\gc) is regarded as being frozen at its vacuum value ($\gc_0$). This approximation was relaxed in a preceding work \cite{arvDprc} within this effective hadronic model, where the in-medium modifications of this dilaton field were found to be quite meager. We conclude, therefore, that this weak dependence only serves to justify the validity of this assumption. 

We next describe our analysis for the in-medium behavior of \dsm s, beginning first with the  nuclear matter $(f_s = 0)$ situation, and including the hyperonic degrees of freedom only later. This approach has the advantage that many features of the \dsm \ in-medium behavior, common between these regimes, are discussed in detail in a more simplified context, and the effect of strangeness becomes a lot clearer.

In nuclear matter, the \wtt \ and the $d_2$ range term vanish, since they depend on the number densities and scalar densities of the hyperons, and have no contribution from the nucleons. It follows from the self-energy expression that only the \wtt \ differs between \dsplus \ and \dsminus; all other interaction terms are absolutely identical for them. Thus, a direct consequence of the vanishing of Weinberg-Tomozawa contribution is that, \dsplus \ and \dsminus \ are degenerate in nuclear matter. As mentioned earlier, the \dsm \ dispersion relations, with $\vec{k} = 0$, eqns.(\ref{disp_a}-\ref{disp_c}), reduce to the quadratic equation form $(A \go^2 + B \go + C = 0)$, with the following coefficients (in nuclear matter):
\begin{eqnarray} 
A_{(N)} & = & \Big[ 1 + \Big( \frac{d_1}{2\fds^2} \big( \grssub{p} 
+ \grssub{n} \big) \Big) -\Big( \frac{\sqrt{2}}{\fds}
\big( \gz' + \gz_c' \big) \Big) \Big] \label{disp_nucl_A}\\
B_{(N)} & = & 0 \\
C_{(N)} & = & \Big[-\mds^2 + 
 \frac{\mds^2}{\sqrt{2}\fds}\big( \gz' + \gz_c' \big) \Big]
\label{disp_nucl}
\end{eqnarray}
where the subscript $(N)$ emphasizes on the nuclear matter context. For solving this quadratic equation, we require the values of $\gz'$, as well as the scalar densities of protons and neutrons, which are obtained from a simultaneous solution of coupled equations of motion for the scalar fields, subject to constraints of fixed values of \grb \ and $\eta$. The behavior of the scalar fields, so obtained, has been discussed in detail in Ref.\cite{arvDprc}.  
Here, we build upon these scalar fields and proceed to discuss the behavior of solutions of the in-medium dispersion relations of \dsm s, given by equations (\ref{disp_nucl_A}) to (\ref{disp_nucl}). 
\begin{figure}
\begin{center}
\scalebox{0.8}{\includegraphics{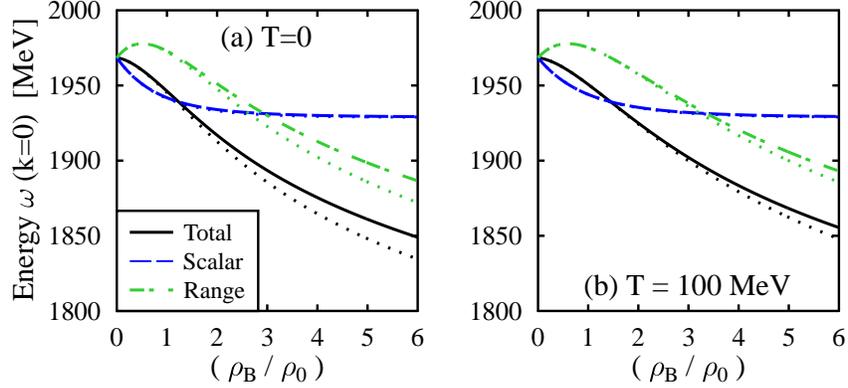}}
\caption{\label{DS_Nucl_TbyT} (Color Online) 
The various contributions to the \dsm \ energies, $\go(\vec{k} = 0)$, in nuclear matter, plotted as functions of baryonic density in units of the nuclear saturation density (\grb/\grz). In each case, the isospin asymmetric situation ($\eta = 0.5$), as described in the legend, is also compared against the symmetric situation ($\eta = 0$), represented by dotted lines.}
\end{center}
\end{figure}
The variation of the \dsm \ in-medium mass, $\go ({\vec k} = 0)$, with baryonic density in nuclear matter, along with the individual contributions to this net variation, is shown in Fig.\ref{DS_Nucl_TbyT} for both zero and a finite temperature value ($T = 100$ MeV). It is observed that the in-medium mass of \dsm s, decreases with density, while being weakly dependent on temperature and \isoap. 
We can understand the observed behavior through the following analysis. 
From equations (\ref{disp_nucl_A}) to (\ref{disp_nucl}), we arrive 
at the following closed form solution for the \dsm \ effective mass 
in nuclear matter: 
\begin{equation}
\go = \mds {\left[ {\frac{1-\frac{\left( \gz' + \gz_c' \right)}{\sqrt{2}\fds}}{1+\frac{d_1}{2\fds^2} \big( \grssub{p} + \grssub{n} \big) -\frac{\sqrt{2}}{\fds}\big( \gz' + \gz_c' \big)}} \right]}^{1/2}
\label{Nucl_exact}
\end{equation}
From this exact closed-form solution, several general conclusions regarding the in-medium behavior of \dsm s in nuclear matter, can be drawn. 
On the basis of this expression, the $d_1$ range term appearing
in the denominator will lead to a decrease in the medium mass 
with increase in density. The in-medium mass, in addition, also
is modified by the medium dependence of $\zeta$ (we assume
the value of $\zeta_c$ to be frozen at its vacuum value
and hence ${\zeta_c}'=0$). 
However, the density dependence of $\gz'$ is observed to be
quite subdued in nuclear matter. It is observed that 
the total change in the value of $\gz$ from its vacuum value, 
i.e. $(\gz')_{\rm max.} \approx 15 \ {\rm  MeV}$. This value 
is acquired at a baryonic density of $\grb \approx 3\grz$, 
and thereafter it appears to saturate. This behavior of the 
strange scalar field, though ubiquitous in this chiral model 
\cite{Pap_prc99,arvDprc}, is not just limited to it. 
Even in calculations employing the relativistic Hartree 
approximation to determine the variation of these scalar fields
 \cite{mam_rha1_2004,mam_rha_2004}, a behavior similar 
to this mean field treatment is observed. 
We also point out that the saturation of the scalar 
meson exchange and 
first range term follows as a direct consequence of the saturation 
of $\gz'$, as is implied by their proportionality in equations 
(\ref{L_int_ds}) and (\ref{disp_nucl}). 
At larger densities, 
while the terms having $\gz'$ saturate, the contribution 
from the $d_1$ range term continues to rise. 
With $\zeta'$ saturating at a value of around 15 MeV
and the value of $f_{D_s}$ chosen to be 235 MeV
in the present investigation,
the last term in the numerator as well as the denominator,  
in equation (\ref{Nucl_exact}) turn out to be much smaller
as compared to unity.
%
Moreover, the $d_1$ term in the denominator in the 
expression for the in-medium mass of $D_s$ given by
equation (\ref{Nucl_exact}) also turns out to be  
much smaller as compared to unity, with 
$(d_1(\rho^s_{p} + \rho^s_{n})/2f_{D_s}^2) \simeq 0.05$ 
at a density of $\rho_B=6\rho_0$. Hence the smallness
of this term as compared to unity is justified 
for the entire range of densities we are concerned with,
in the present investigation.
In order to read more into our solution, we expand the argument 
of the square root, in a binomial series (assuming
$\;\;\frac{\left( \gz' + \gz_c' \right)}{\sqrt{2}\fds} << 1\;\;\; 
{\rm {and}} \;\;\;(d_1(\rho^s_{p} + \rho^s_{n})/2f_{D_s}^2)<<1$), 
and retain up to only 
the first order terms. This gives, as the approximate solution:
\begin{equation}
\go \approx \mds \Big[ 1 + \frac{\left( \gz' + \gz_c' \right)}{2\sqrt{2}\fds} - \frac{d_1}{4\fds^2} \big( \grssub{p} + \grssub{n} \big) \Big]
\label{nucl_binomial}
\end{equation}
Moreover, since \grssub{i} $\approx$ \grsub{i} \ at small densities, 
the $d_1$ range term contribution in the above equation approximately 
equals $(d_1/4\fds^2) \grb$ \ at low densities. 
The first of the terms in the approximate expression given by Eq. (\ref{nucl_binomial}) is an increasing function while the second one is a decreasing function of density. Their interplay generates the observed curve shape - the repulsive contribution being responsible for the small hump in an otherwise linear fall, at small densities ($0<\grb<0.5\grz$). While the attractive $d_1$ term would have produced a linear decrease right away, the role of repulsive contribution is to impede this decrease, hence producing the hump. At moderately higher densities, however, the contribution from the second term outweighs the first, which is why we see a linear drop with density. This is observed to be the case, till around $\grb \approx 2\grz$. At still larger densities, the approximation (\grssub{i} $\approx$ \grsub{i}) breaks down, though our binomial expansion is still valid. Since scalar density falls slower than number density, the term $[-(d_1/4\fds^2)(\grssub{p}+\grssub{n})]$ will fall faster than $[-(d_1/4\fds^2)\ \grb]$. This is responsible for the change in slope of the curve at intermediate densities, where a linear fall with density is no longer obeyed. So, at intermediate and large densities, the manner of the variation is dictated by the scalar densities, in the $d_1$ range term. 
From equations (\ref{disp_a}) to (\ref{disp_c}), one would expect the mass modifications of \dsm s to be insensitive to isospin asymmetry, since, e.g. in this nuclear matter case, the dispersion relation bears isospin-symmetric terms like (\grssub{p}+\grssub{n}). (This is in stark contrast with earlier treatments of kaons and antikaons \cite{sambuddha1,sambuddha2} as well as the \dm s \cite{arindam} within this 
effective model, where the dispersion relations had terms like (\grssub{p}$-$\grssub{n}) or (\grsub{p}$-$\grsub{n}), which contributed to asymmetry.) However, the \dsm \ effective mass is observed to depend on  asymmetry in Fig.\ref{DS_Nucl_TbyT}, though the dependence is weak. For example, the values of $\go ({\vec k} = 0)$, for the isospin symmetric situation $(\eta = 0)$, are $1913$, $1865$ and $1834$ MeV, respectively, at $\grb = 2\grz, \ 4\grz \ {\rm and} \ 6\grz$ while the corresponding numbers for the (completely) asymmetric $(\eta = 0.5)$ situation are $1917$, $1875$ and $1849$ MeV, respectively, at $T=0$. Since the $\eta$-dependence of $\llags{\rm SME}$ and $\llags{\rm 1^{st} Range}$ contributions must be identical, one can reason from Fig.\ref{DS_Nucl_TbyT} that this isospin dependence of \dsm \ effective mass is almost entirely due to the $d_1$ range term ($\sim$ (\grssub{p} + \grssub{n})), which was expected to be isospin symmetric. This apparently counterintuitive behavior has been observed earlier in \cite{arvDprc}, in the context of \dm s. This is because, the value of (\grssub{p} + \grssub{n}) turns out to be different for symmetric and asymmetric situations, contrary to naive expectations. Since the scalar-isovector \gd \ meson ($\gd \sim \langle{\bar u}u-{\bar d}d\rangle$) is responsible for introducing isospin asymmetry in this effective hadronic model \cite{Pap_prc99}, owing to the equations of motion of the scalar fields being coupled, the values of the other scalar fields turn out to be different in the symmetric $(\gd = 0)$ and asymmetric $(\gd \ne 0)$ cases. The same is also reflected in the values of the scalar densities calculated from these scalar fields, which leads to the observed behavior. 

Additionally, we observe from a comparison of figures \ref{DS_Nucl_TbyT}a 
and \ref{DS_Nucl_TbyT}b that the magnitude of the \dsm \ mass drop decreases 
with an increase in  temperature from T=0 to T=100 MeV. 
For example, in the symmetric ($\eta = 0$) 
situation, at $T=0$, the \dsm \ mass values, at $\grb = 2\grz, \ 4\grz \ 
{\rm and} \ 6 \grz$, are $1913, \ 1865 \ {\rm and} \ 1834$ MeV respectively, 
which grow to $1925, \ 1879 \ {\rm and} \ 1848$ MeV respectively, 
at $T=100$ MeV. Likewise, with $\eta = 0.5$, the corresponding values 
read $1917, \ 1875 \ {\rm and} \ 1849$ MeV at $T = 0$, while the same 
numbers, at $T=100$ MeV, are $1925, \ 1883 \ {\rm and} \ 1855$ MeV. 
Thus, though small, there is a definite reduction in the magnitude 
of the mass drops, as we go from $T=0$ to $T=100$ MeV, in each 
of these cases. This behavior can be understood, from the point 
of view of the temperature variation of scalar condensates, 
in the following manner. It is observed that the scalar fields 
decrease with an increase in temperature from $T=0$ to $100$ MeV 
\cite{arvDprc,mamD2004}. In Ref.\cite{mamD2004}, the same effect 
was understood as an increase in the nucleon mass with temperature. 
The equity of the two arguments can be seen by invoking the expression 
relating the baryon mass to the scalar fields' magnitude, 
$m_i^* = -(g_{\gs i} \gs + g_{\gz i} \gz + g_{\gd i} \gd)$ \cite{Pap_prc99}. 
The temperature dependence of the scalar fields ($\sigma$, $\zeta$,
$\delta$) have been studied within the model in \cite{arvDprc},
which are observed to be different for the zero and finite baryon
densities. At zero baryon density, the magnitudes of the scalar 
fields $\sigma$ and $\zeta$ are observed to remain almost constant 
upto a temperature of about 125 MeV, above which these are observed
to decrease with temperature. This behaviour can be
understood from the 
expression of $\rho_s^i$ 
given by equation 
(\ref{scalardensityintegral}) for the situation of zero density,
i.e. for $\mu_i^*$=0, which decreases with increase with temperature. 
The temperature dependence of the scalar density in turn 
determines the behaviour of the scalar fields. The scalar fields
which are solved from their equations of motion,
behave in a similar manner as the scalar density. 
At finite densities,
i.e., for nonzero values of the effective chemical potential, 
$\mu_i^*$, however, the temperature dependence
of the scalar density is quite different from the zero density
situation. For finite baryon density, with increase in temperature, 
there are contributions 
also from higher momenta, thereby increasing the denominator of the
integrand on the right hand side of the baryon
scalar density given by Eq. (\ref{scalardensityintegral}). 
This leads to a decrease in the scalar density. At finite
baryon densities, the competing effects of
the thermal distribution functions and the contributions from
the higher momentum states give rise to the temperature 
dependence of the scalar density, which in turn determine the
behaviour of the $\sigma$ and $\zeta$ fields with temperature.
These scalar fields are observed to have
an initial increase in their magnitudes 
upto a temperature of around 125-150 MeV, followed by a 
decrease as the temperature is further raised \cite{arvDprc}.
This kind of behavior of the
scalar $\sigma$ field on temperature at finite densities has also been
observed in the Walecka model in Ref. \cite{LiKoBrown}.
In fact, we point out that a decrease in the scalar condensates with 
an increase in temperature, though small in the hadronic regime 
($<170$ MeV), is well-known in general model-independent terms 
\cite{Gerber_Leutwyler_NPB_1989}, and was also observed to be 
a consistent feature of all $U(N_f)_L \times U(N_f)_R$ linear sigma 
models in the model-independent work of R\"oder et al. \cite{roder}. 
Since these scalar fields serve as an input in calculating the scalar 
densities \cite{Pap_prc99}, a decrease in the magnitude of the latter 
accompanies a decrease in the former, at larger temperatures. 
From the point of view of the dispersion relations, this results 
in a decrease in the coefficient $A$, and owing to the inverse 
dependence of $\go$ on $A$, increases the value of $\go({\vec k = 0})$ 
in the finite temperature case, as compared to the $T=0$ situation. 
Or stating it differently, the difference of this $\go({\vec k = 0})$, 
from the vacuum value, 
i.e the mass drop, decreases. 
Thus, from a physical viewpoint, since the origin of these mass drops 
is the attractive in-medium interactions, one can say that the reduction 
in the mass drop magnitudes is due to a weakening of the attractive 
strength of these in-medium interactions, represented in these models 
by a reduction in the quark condensates with increasing temperatures. 
Additionally, it is also observed from Fig.\ref{DS_Nucl_TbyT} that isospin dependence of the \dsm \ mass, feeble anyways, weakens further with temperature. For example, as mentioned earlier, mass of the \dsm s at $\grb = 6 \grz$, for the $\eta = 0$ and $\eta = 0.5$ cases, are $1834$ and $1849$ MeV respectively (a difference of $15$ MeV), when $T=0$, which changes to $1848$ and $1855$ MeV at $T = 100$ MeV (a $7$ MeV difference). This is, once again, due to a decrease in the magnitude of \gd, with temperature, at any fixed value of the parameters $\eta$ and \grb. In particular, the difference between the value of \gd \  for the $\eta = 0$ and $\eta = 0.5$ cases is observed to decrease with temperature (as was shown explicitly in Ref. \cite{arvDprc}). Since asymmetry is introduced through \gd, a decrease in the difference of the values of $\go$, between symmetric and asymmetric situations, with temperature, follows naturally.

Next, we generalize our analysis by including hyperonic degrees of freedom as well, in the medium. However, as mentioned previously, in the ensuing discussion of \dsm s in hyperonic matter, we focus predominantly on the new physics arising via the introduction of strangeness in the medium, since, e.g. the weak dependence on isospin asymmetry, or a weak reduction in the mass drop magnitudes at higher temperatures, has, in principle, the same explanation that stood in the nuclear matter context. 
\begin{figure}
\begin{center}
\scalebox{0.8}{\includegraphics{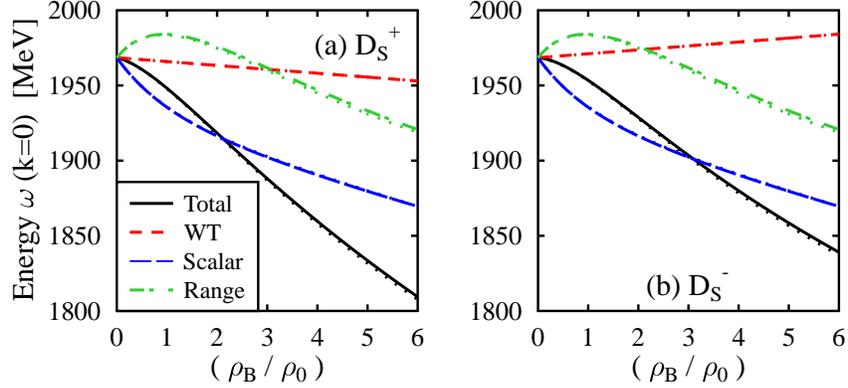}}
\caption{\label{Hyp_TbyT} (Color Online) The various contributions to the effective mass of \dsplus \ and \dsminus \ mesons, as a function of baryonic density, for typical values of temperature ($T=100$ MeV) and strangeness fraction ($f_s = 0.5$). In each case, the asymmetric situation ($\eta = 0.5$), as described in the legend, is also compared against the symmetric situation ($\eta = 0$), represented by dotted lines.}
\end{center}
\end{figure}

Fig.\ref{Hyp_TbyT} shows the variation of the in-medium mass of the \dsm s, along with the various individual contributions, in hyperonic medium as a function of baryonic density, for typical values of temperature, \isoap \ and strangeness fraction. The most drastic consequence of the inclusion of hyperons in the medium is that, the mass degeneracy of \dsplus \ and \dsminus \ now stands 
broken. For example, the values of $(m_{\dsplus}, m_{\dsminus})$ at $\grb = \grz, 2\grz, 4\grz \ {\rm and} \ 6\grz$, are observed to be $(1948,1953)$, $(1918,1928)$, $(1859,1879)$ and $(1808,1838)$ MeV, respectively, as can be seen from the figure.
Thus, except at vacuum ($\grb = 0$), mass difference of \dsplus \ and \dsminus \ is non-zero at finite \grb, growing in magnitude with density. This mass degeneracy breaking is a direct consequence of non-zero contributions from the \wtt, which follows from equations (\ref{L_int_ds}), (\ref{selfenergy}) and (\ref{disp_b}). The same may be reconciled with Fig.\ref{Hyp_TbyT}, from which it follows that except for this \wtt \ acquiring equal and opposite values for these mesons, all other terms are absolutely identical for them.
Once again, on the basis of the following analysis of  the \dsm \ dispersion relations at zero momentum, we insist that this observed behavior is perfectly consistent with expectations.

\begin{figure}
\begin{center}
\scalebox{0.8}{\includegraphics{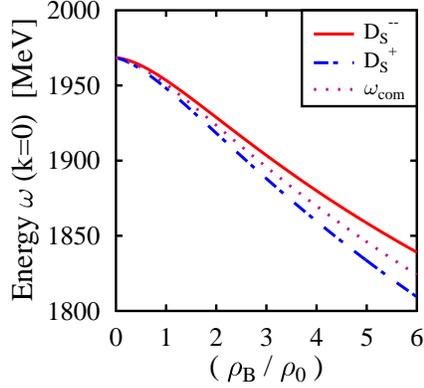}}
\caption{\label{Hyp_sym_fall} (Color Online) \dsm \ mass degeneracy breaking in hyperonic matter, as a function of baryonic density, for typical values of the other parameters ($T=100$ MeV, $f_s = 0.5$ and $\eta = 0.5$). The medium masses of \dsplus \ and \dsminus \ are observed to fall symmetrically about $\go_{\rm com}$ (\emph{see text}).}
\end{center}
\end{figure}
The general solution of the equivalent quadratic equation, 
eqn.(\ref{disp_relatn_k0_quad}), is:
\begin{eqnarray}
\go = \frac{(-B + \sqrt{B^2 - 4AC})}{2A} \approx  -\frac{B}{2A} 
+ {\sqrt {\frac{C_1}{A}}} + \frac{B^2}{8A\sqrt{AC_1}} + \ldots
\end{eqnarray}
where we have disregarded the negative root, and have performed a binomial expansion of 
$(1 + B^2/4AC_1)^{1/2}$, with $C_1 = -C$. Upon feeding numerical values, we observe that the expansion parameter $(B^2/4AC_1)$ is much smaller than unity for our entire density variation, which justifies the validity of this expansion for our analysis. The same exercise also allows us to safely disregard higher-order terms, and simply write:
\begin{equation}
\go_{\rm hyp} \approx {\sqrt {\frac{C_1}{A}}}  -\frac{B}{2A} 
\label{sign_reversal_eqn}
\end{equation}
Further, with the same justification as in the nuclear matter case, both $(C_1/A)^{1/2}$ and $(1/A)$ can also be expanded binomially. For example, for the second term, this gives 
\begin{equation}
\frac{|B|}{2A} = \frac{|B|}{2} + {\cal{O}}\Big (
\frac{\grsub{i}\grssub{i}}{\fds^{\ 4}}\Big)
\end{equation}
where the contribution from these higher order terms is smaller owing to the large denominator, prompting us to retain only the first order terms. Here, we point out that since $B = \pm |B|$, ($+$ sign for \dsplus \ and $-$ sign for \dsminus), this term, which represents the Weinberg-Tomozawa contribution to the dispersion relations, breaks the degeneracy of the \dsm s. On the other hand, the first term in eqn.(\ref{sign_reversal_eqn}) is common between \dsplus \ and \dsminus \ mesons. Thus, the general solution of the \dsm \ dispersion relations in the hyperonic matter context, can be written as: 
\begin{equation}
\go_{\rm hyp} = \go_{\rm com} \mp \go_{\rm brk},
\end{equation} 
where 
\begin{eqnarray}
\go_{\rm com} & = & {\sqrt {\frac{C_1}{A}}} \approx \mds 
\big[1 - \Big( \frac{d_1}{4\fds^2} {\sum_{(N+H)}} \grssub{i} \Big) 
\big.\nonumber\\ 
& & - \Big. \Big( \frac{d_2}{4\fds^2} \Big( 2 {\sum_{\Xi}} \grssub{i} 
+ {\sum_{(H-\Xi)}}\grssub{i} \Big) \Big) +\Big( \frac{\Big( \gz' 
+ \gz_c' \Big)}{2\sqrt{2}\fds} \Big) \big],
\label{commonpartofthesolution}
\end{eqnarray}
\begin{equation}
\go_{\rm brk} = \frac{|B|}{2A} \approx \frac{|B|}{2} = \Big[ 
\frac{1}{4\fds^2} \Big(  2 {\sum_{\Xi}} \grsub{i} \ 
+ {\sum_{(H-\Xi)}} \grsub{i} \Big) \Big]
\label{breakingpartofthesolution}
\end{equation}

Thus, $\go_{\dsplus} = \go_{\rm com} - \go_{\rm brk}$, and $\go_{\dsminus} = \go_{\rm com} + \go_{\rm brk}$, (where $\go_{\rm brk}$ is necessarily positive), which readily explains why the mass of \dsplus \ drops more than that of \dsminus, with density.   Also, this formulation readily accounts for symmetric fall of the medium masses of \dsplus \ and \dsminus \ mesons, about $\go_{\rm com}$ in Fig.\ref{Hyp_sym_fall}. Trivially, it may also be deduced that $\go_{\rm brk}$ is extinguished in nuclear matter, so that the mass degeneracy of \dsplus \ and \dsminus \ is recovered from these equations. However, we observe that though the curve corresponding to $\go_{\rm com}$ is exactly bisecting the masses of \dsplus \ and \dsminus \ at small densities, this bisection is no longer perfect at high densities. This can be understood in the following manner. In essence, we are comparing the density dependence of a function $f(\grssub{i})$, with the functions  $f(\grssub{i}) \pm g(\grsub{i})$. Since scalar densities fall sub-linearly with the number density, it follows that the fall can not be absolutely symmetric at any arbitrary density. In fact, since a decreasing function of scalar density will fall faster than that of a number density, one expects $\go_{\rm com}$ to lean away from the curve for \dsminus \ meson and towards the curve corresponding to \dsplus, which is exactly what is observed in Fig.\ref{Hyp_sym_fall}.

\begin{figure}
\begin{center}
\scalebox{0.8}{\includegraphics{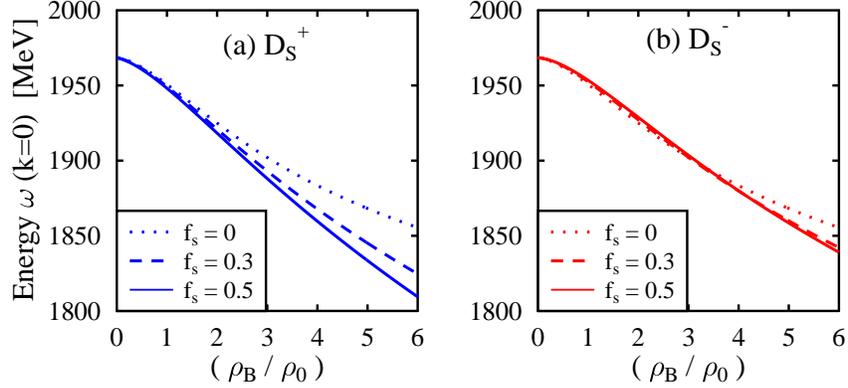}}
\caption{\label{DSplusminus_fsvary} (Color Online) The sensitivity of the medium mass of (a) \dsplus, and (b) \dsminus \ mesons, to the strangeness fraction $f_s$, at typical values of temperature ($T=100$ MeV) and \isoap \ ($\eta = 0.5$).}
\end{center}
\end{figure}
Since this disparity between \dsplus \ and \dsminus \ originates from the \wtt, whose magnitude is directly proportional to the hyperonic number densities, one expects this disparity to grow with an increase in the strangeness content of the medium. The same also follows from the above formulation, since the mass splitting between the two, $\Delta m = (m_{\dsminus} - m_{\dsplus}) \equiv ({\go}_{\dsminus} - {\go}_{\dsplus}) = -2{\go}_{\rm brk}$, should grow with both, $f_s$ at fixed \grb \ (i.e. a larger proportion of hyperons), as well as with \grb \ at fixed (non-zero) $f_s$ (i.e. a larger hyperonic density), in accordance with eqn.(\ref{breakingpartofthesolution}). Naively, one expects this $f_s$ dependence to be shared by the two \dsm s; however, closer analysis reveals that this is not the case, as shown in Fig.\ref{DSplusminus_fsvary} where we consider the $f_s$ dependence of the \dsm \ medium mass. It is observed that while the $f_s$ dependence for \dsplus \ meson is quite pronounced, the same for the \dsminus \ is conspicuously subdued. Counterintuitive as it may apparently be, this observed behavior follows from the above formulation. Since $\go_{\dsminus} = \go_{\rm com} + \go_{\rm brk}$, at any fixed density, the first of these is a decreasing function of $f_s$, while the second increases with $f_s$. These opposite tendencies are responsible for the weak $f_s$ dependence of \dsminus \ meson. On the other hand, because $\go_{\dsplus} = \go_{\rm com} - \go_{\rm brk}$, the two effects add up to produce a heightened overall decrease with $f_s$ for the \dsplus \ meson, as we observe in the figure.

\begin{figure}
\begin{center} 
\scalebox{0.8}{\includegraphics{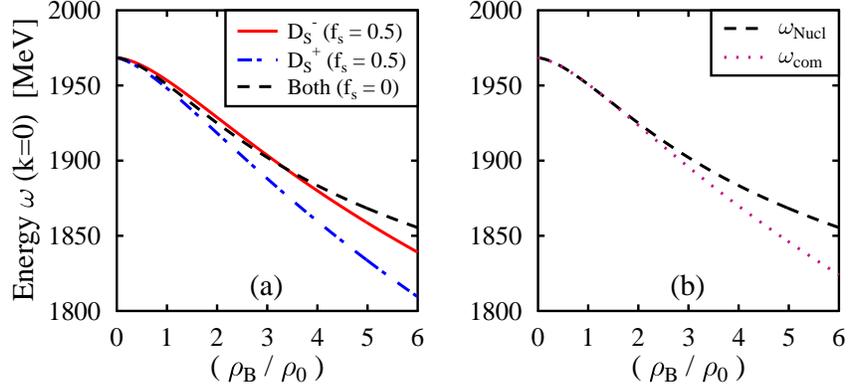}}
\caption{\label{Hyp_vs_nucl} (Color Online) (a) A comparison of medium mass of \dsm s, in nuclear and hyperonic matter, maintaining the same values of parameters as before. (b) Comparison of the nuclear matter solution, $\go_{\rm nucl}$, with the $\go_{\rm com}$ in hyperonic matter, as observed in Fig.\ref{Hyp_sym_fall}.}
\end{center} 
\end{figure}

A comparison of these hyperonic matter $(f_s \ne 0)$ solutions with the nuclear matter $(f_s = 0)$ solutions is shown in Fig.\ref{Hyp_vs_nucl} for typical values of the parameters. It is observed from Fig.\ref{Hyp_vs_nucl}a that at small densities, the nuclear matter curve, which represents both \dsplus  \ and \dsminus, bisects the mass degeneracy breaking curve, akin to $\go_{\rm com}$ in Fig.\ref{Hyp_sym_fall}. However, at larger densities, both these $f_s \ne 0$ curves drop further than the $f_s = 0$ curve. From a casual comparison of expressions, it appears that the nuclear matter solution, given by eqn.(\ref{nucl_binomial}), also enters the hyperonic matter solution, eqn.(\ref{commonpartofthesolution}), albeit as a subset of $\go_{\rm com}$. It is tempting to rearrange the latter then, such that this nuclear matter part is separated out, generating an expression of the type $\go_{\rm hyp} = \go_{\rm nucl} + f(\grsub{i},\grssub{i})|_{i = H},$ which would also entail the segregation of the entire $f_s$ dependence. However, careful analysis reveals that it is impossible to achieve this kind of a relation since, in spite of an identical expression, in hyperonic matter, the RHS of eqn.(\ref{nucl_binomial}) does not give the nuclear matter solution. This is because, the nuclear matter solution involves the scalar fields (and scalar densities computed using these scalar fields) obtained as a solution of coupled equations in the $f_s = 0$ situation, which are radically different from the solutions obtained in the $f_s \ne 0$ case. (This has been observed to be the case, in almost every treatment of hyperonic matter within this 
effective model, but perhaps, most explicitly in \cite{arvDepja}.) Thus, one can not retrieve the nuclear matter solution from the scalar fields obtained as solutions in the $f_s \ne 0$ case; moreover, due to the complexity of the concerned equations, it is impossible to achieve a closed-form relation between the values of the scalar fields (and hence, also the scalar densities) obtained in the two cases. In fact, the low-density similarity between 
Fig.\ref{Hyp_sym_fall} and Fig.\ref{Hyp_vs_nucl}a might lead one to presume that $\go_{\rm com}$ reduces to $\go_{\rm nucl}$ at small densities. This expectation is fueled further by their comparison in Fig.\ref{Hyp_vs_nucl}b, which shows clearly that they coincide at small densities. However, in light of the above argument, it follows that they are absolutely unrelated. Their coincidence at small densities can be explained by invoking the relation $\grssub{i} \approx \grsub{i}$ in this regime, using which, eqn.(\ref{commonpartofthesolution}) reduces to  
\begin{equation}
\go_{\rm com} \approx \mds \Big( 1 + \frac{\left( \gz' + \gz_c' 
\right)}{2\sqrt{2}\fds} - \frac{\kappa_1}{4\fds^2} \grb \Big) - \mds \Big(
 \frac{\kappa_2 \Delta}{4\fds^2} \Big), 
\label{label_blablabla}
\end{equation}
with $\kappa_1 = (d_1 + d_2) \approx 1.28 d_1$, with our choice of parameters, $\kappa_2  \ (\equiv d_2) = 0.22 \kappa_1$ and the difference term, $\Delta = (\grssub{\gX^0}+\grssub{\gX^-}-\grssub{p}-\grssub{n})$. The contribution from this second term in eqn.(\ref{label_blablabla}) is significantly smaller than the first term at small densities, owing to the smaller coefficient, as well as the fact that this depends on the difference of densities rather than on their sum (like the first term). This first part of eqn.(\ref{label_blablabla}) can be likened to the approximate nuclear matter solution at small \grb, as we concluded from eqn.(\ref{nucl_binomial}). The marginal increase of $\kappa$ above $d_1$ is compensated by a marginal increase in $\gz'$ for the $f_s \ne 0$ case, as compared to the $f_s = 0$ situation, and hence, the two  curves look approximately the same at small densities in Fig.\ref{Hyp_vs_nucl}b. At larger densities however, this simple picture breaks down, and the fact that they are unrelated becomes evident. Nevertheless, we may conclude in general from this comparison that the inclusion of hyperonic degrees of freedom in the hadronic medium makes it more attractive, regarding \dsm s' in-medium interactions, especially at large \grb. From the point of view of the \dsm \ dispersion relations, this is conclusively because of the contributions from the extra, hyperonic terms, which result in an overall increase of the coefficient $A_{\rm hyp}$ significantly above $A_{\rm nucl}$, particularly at large $f_s$.

\begin{figure}
\begin{center}
\scalebox{0.8}{\includegraphics{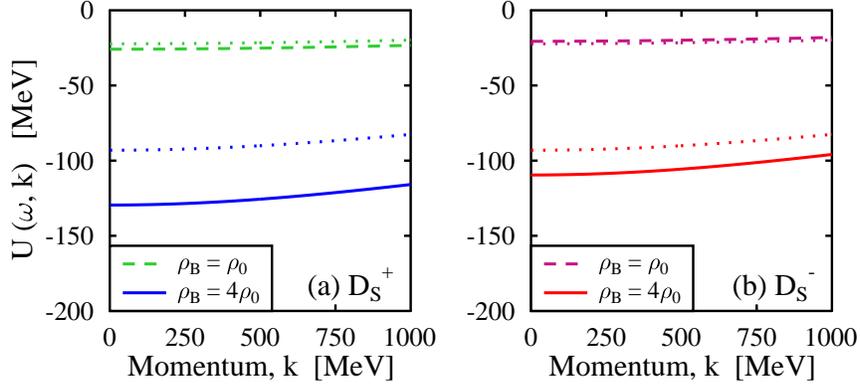}}
\caption{\label{OptPot} (Color Online) The optical potentials of the, (a) \dsplus \ and (b) \dsminus \ mesons, as a function of momentum $k \ (\equiv |{\vec k}|)$, in cold $(T=0)$, asymmetric ($\eta = 0.5$) matter, at different values of \grb. 
In each case, the hyperonic matter situation ($f_s = 0.5$), as described in the legend, is also compared against the nuclear matter ($f_s = 0$) situation, represented by dotted lines.}
\end{center}
\end{figure}

Finally, we consider momentum-dependent effects by investigating the behavior of the in-medium optical potentials for the \dsm s. These are shown in Fig.\ref{OptPot}, where we consider them in the context of both, asymmetric nuclear ($f_s = 0$) as well as hyperonic matter ($f_s = 0.5$), as a function of momentum $k \ (= \vert{\vec k}\vert)$ at fixed, values of the parameters \grb, $\eta $ and $T$. As with the rest of the investigation, the dependence of the in-medium optical potentials on the parameters $T$ and $\eta$ is quite weak; hence, we neglect their variation in this context. In order to appreciate the observed behavior of these optical potentials, we observe that as per its definition, eqn.(\ref{OptPot_Def}), at zero momentum, optical potential is just the negative of the mass drop of the respective meson (i.e. $U(k=0) = - \Delta m (k=0) \equiv \Delta m (\grb, T, \eta, f_s) )$. It follows then, that at $k=0$, the two \dsm s are degenerate in nuclear matter; moreover, it is observed that this degeneracy extends to the finite momentum regime. This is because, from equations (\ref{dispersion}) and (\ref{selfenergy}), the finite momentum contribution is also common for \dsplus \ and \dsminus \ in nuclear matter. In hyperonic matter, however, non-zero contribution from these terms breaks the degeneracy in the $k=0$ limit (as was already discussed before), and finite momentum preserves this non-degeneracy. Consequently, at any fixed density, the curves for different values of $f_s$ differ in terms of their y-intercept but otherwise, appear to run parallel. Further, as we had reasoned earlier for the $k=0$ case, the effect of increasing $f_s$ on the mass drops of \dsplus \ and \dsminus \ is equal and opposite about the nuclear matter situation at small densities, which readily explains the behavior of optical potentials for these mesons at $\grb = \grz$. At larger densities, e.g. the $\grb = 4\grz$ case shown in Fig.\ref{OptPot}, the lower values of optical potentials for both \dsplus \ and \dsminus, as compared to the $f_s = 0$ case, can be immediately reconciled with the behavior observed in Fig.\ref{Hyp_vs_nucl}. In fact, the large difference between the \dsplus \ optical potential for the $f_s = 0$ and $f_s = 0.5$ cases, as compared to that for \dsminus, is a by-product of their zero-momentum behavior, preserved at non-zero momenta as a consequence of equations (\ref{dispersion}) and (\ref{selfenergy}).

We next discuss the possible implications of these medium effects. 
In the present investigation, we have observed a reduction in the mass of the \dsm s, with an increase in baryonic density. This decrease in \dsm \ mass can result in the opening up of extra reaction channels of the type $A \rightarrow \ \dsplus \dsminus$ in the hadronic medium.  
\begin{figure}
\begin{center}
\scalebox{0.9}{\includegraphics{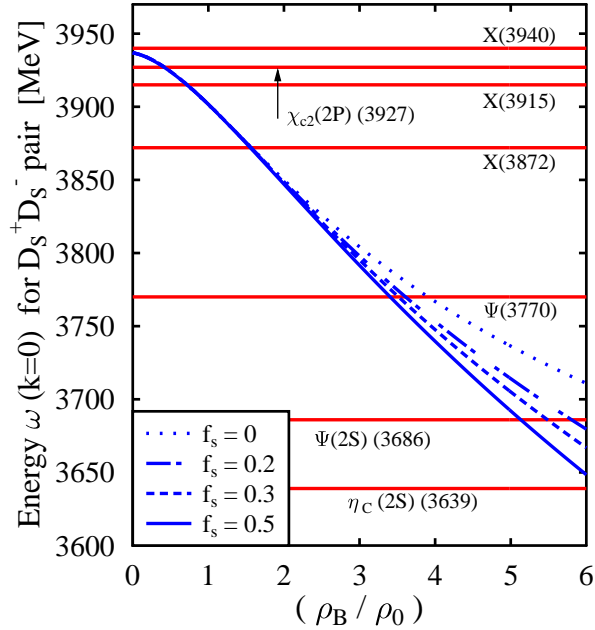}}
\caption{\label{Decays} (Color Online) Mass of the \dsplus\dsminus \ pair, compared against the (vacuum) masses of the excited charmonia states, at typical values of temperature ($T=100$ MeV) and asymmetry ($\eta = 0.5$), for both nuclear matter ($f_s = 0$) and hyperonic matter ($f_s = 0.2, \ 0.3 {\rm \ and \ } 0.5$) situations. The respective threshold density (\emph{see text}) is given by the point of intersection of the two concerned curves.}
\end{center}
\end{figure}
In fact, a comparison with the spectrum of 
known charmonium states \cite{PDG2012,Voloshin} sheds light on the possible decay channels, as shown in Fig.\ref{Decays}. As a starting approximation, we have neglected the variation of energy levels of these charmonia with density, since we intuitively expect medium effects to be larger for an $s{\bar c}$ (or $c{\bar s}$) system, as compared to a $c{\bar c}$ system. However, this approximation can be conveniently relaxed, e.g. as was done in Refs. \cite{arvDepja, DP_Bmonia_PRC} for charmonia and bottomonia states, respectively. It  follows from Fig.\ref{Decays} that above certain threshold density, which is given by the intersection of the \dsm \ curve with the respective energy level, the mass of the \dsplus \dsminus \ pair is smaller than that of the excited charmonium state. Hence, this decay channel opens up above this threshold density. The most immediate experimental consequence of this effect would be a decrease in the production yields for these excited charmonia in collision experiments. Additionally, since an extra decay mode will lower the lifetime of the state, one expects the decay widths of these excited charmonia to be modified as a result of these extra channels. Though it can be reasoned right away that reduction in lifetime would cause a broadening of the resonance curve, more exotic effects can also result from these level crossings. For example, it has been suggested in Ref. \cite{friman}, that if the internal structure of the hadrons is also accounted for, there is even a possibility for these in-medium widths to become narrow. This counterintuitive result originates from the node structure of the charmonium wavefunction, which can even create a possibility for the decay widths to vanish completely at certain momenta of the daughter particles. Also, if the medium modifications of the charmonium states are accounted for, the decay widths will be modified accordingly (as was observed in the context of the \dm s, in \cite{arvDepja}). 

Apart from this, one expects signatures of these medium effects to be reflected in the particle ratio ($\dsplus/\dsminus$). 
A production asymmetry between \dsplus \ and \dsminus \ mesons 
might be anticipated owing to their unequal medium masses, 
as observed in the current investigation. 
Also, due to the semi-leptonic and leptonic decay modes of the \dsm s 
\cite{PDG2012}, one also expects that, accompanying an increased production 
of \dsm s due to a lowered mass, there would be an enhancement in dilepton 
production as well. Further, we expect these medium-modifications of 
the \dsm s to be reflected in the observed dilepton spectra, due to 
the well-known fact that dileptons, with their small interaction cross 
section in hadronic matter, can serve as probes of medium effects 
in collision experiments \cite{Vogt_book, li_ko}.  
The attractive interaction of the $D_s$ mesons with the hadronic
medium might also lead to the possibility of formation of 
$D_s$-nucleus bound state, which can be explored at the CBM experiment 
at FAIR in the future facility at GSI \cite{CBM_Physics_Book}. 

We now discuss how the results of the present 
investigation compare with the available literature on the medium 
effects for pseudoscalar $D_S(1968.5)$ mesons.
As has already been mentioned, 
Refs. \cite{lutzkorpa, hofmannlutz, JimenezTejero_2009vq, JimenezTejero_2011fc}
have treated the medium behavior of \dsm s, using the
coupled channel framework. The broad perspective that emerges 
from all these analyses is that of an attractive interaction 
in the medium \cite{JimenezTejero_2009vq, JimenezTejero_2011fc} 
between \dsm s and baryon species like the nucleons or 
the $\gL, \gS, \gX$ hyperons 
\cite{lutzkorpa, hofmannlutz, JimenezTejero_2009vq, JimenezTejero_2011fc}, 
which is consistent with what we have observed in this work. 
Additionally, one observes in these approaches an attractive medium 
interaction even with the charmed baryons \cite{hofmannlutz}. 
Some of these interaction channels also feature 
resonances, which is reflected in the relevant scattering amplitudes 
picking up imaginary parts as well, e.g. the $N_{{\bar c}s}(2892)$ 
state in the $\dsplus N$ channel \cite{lutzkorpa, hofmannlutz}. 
Treating such resonances comes naturally to the coupled channel framework, 
since by default, the strategy is aimed at considering the scattering 
of various hadron species off one another, and assessing scattering 
amplitudes and cross sections etc. \cite{Oset_PRL98, Oset_NPA98}. 
In the present work, we have investigated the in-medium
masses of the $D_s$ mesons arising due to their interactions with the
baryons and scalar mesons in the nuclear (hyperonic) medium,
and have not studied the decays of these mesons. 
As has already been mentioned the mass modifications
of the $D_s$ mesons can lead to opening up of the
new channels for the charmonium states decaying 
to $D_s^+ {D_s^-}$ at certain densities as can be
seen from Fig. \ref{Decays}.
In fact, even if one contemplates working along the lines of 
Ref. \cite{arvDepja} for calculating these decay widths,
the situation is a bit more complicated 
for these \dsm s, since the states 
$X(3872), X(3915)$ and $X(3940)$, which are highly relevant 
in this context (as one can see from Fig. \ref{Decays}), 
and are not classified as having clear, definite quantum numbers 
in the spectrum of excited charmonium states \cite{PDG2012}. 
This makes the assessment of medium effects for these states 
more difficult in comparison, since the identification of 
the relevant charmonium states with definite quantum numbers 
in the charmonium spectrum, such as $1S$ for $J/\psi$, $2S$ 
for $\psi(3686)$ and $1D$ for $\psi(3770)$, was a crucial 
requirement for evaluating their mass shifts in the medium 
in the QCD second-order Stark effect study in Ref. \cite{arvDepja}. 
For some of these unconventional states, e.g. $X(3872)$, this absence 
of definite identification with states in the spectrum, has led 
to alternative possibilities being explored for these states. 
A prominent example is the possibility of a molecular structure 
for this $X(3872)$ state \cite{Voloshin}, borne out of contributions 
from the $( D^+ D^{\ast -} + D^- D^{\ast +})$ and $( D^0 {\bar D}^{\ast 0} 
+ {\bar D}^0 D^{\ast 0})$ components in the $S-$wave. 

Lastly, we point out that the medium effects described in this article, and 
the possible experimental consequences entailed by these, 
are especially interesting in wake of the upcoming 
CBM \cite{CBM_Physics_Book} 
experiment at FAIR, GSI, where high baryonic densities are expected 
to be reached. Due to the strong density dependence of these medium 
effects, we expect that each of these mentioned experimental 
consequences would intensify at higher densities, and should 
be palpable in the aforementioned future experiment. 

\end{section}

\begin{section}{Summary} 

To summarize, we have explored the properties of \dsm s in a hot 
and dense hadronic environment, within the framework 
of the Chiral \chigrp \ model, generalized to $SU(4)$.
The generalization of chiral SU(3) model to SU(4) is done 
in order to derive the interactions of the charmed mesons
with the light hadrons, needed for the study of the
in-medium properties of the $D_s$ mesons in the present
work. However, realizing that the chiral symmetry is 
badly broken for the SU(4) case due to the large mass of the
charm quark, we use the interactions derived from SU(4)
for the $D_s$ meson, but use 
the observed masses of these heavy pesudoscalar mesons
as well as empirical/observed values of their decay constants 
\cite{liukolin}.  
The $D_s$ mesons have been considered in both (symmetric and asymmetric) 
nuclear and hyperonic matter, at finite densities that extend 
slightly beyond what can be achieved with the existing and 
known future facilities, and at both zero and 
finite temperatures. Due to net attractive interactions in the medium, 
\dsm s are observed to undergo a drop in their effective mass. 
These mass drops are found to intensify with an increase in the 
baryonic density of the medium, while being largely insensitive 
to changes in temperature as well as the \isoap. However, 
upon adding hyperonic degrees of freedom, the mass degeneracy 
of \dsplus \ and \dsminus \ is observed to be broken. The mass 
splitting between \dsplus \ and \dsminus \ is found to grow 
significantly with an increase in baryonic density as well as 
the strangeness content of the medium. Through a 
detailed analysis of the in-medium dispersion relations 
for the \dsm s, we have shown that the observed behavior, 
follows precisely from the interplay of contributions 
from various interaction terms in the \lagd. 
We have briefly discussed the possible experimental 
consequences of these medium effects, e.g., in the
$D_s^+/D_s^-$ ratio, dilepton spectra, possibility 
of formation of exotic $D_s$-nucleus bound states,
as well as modifications of the decays of charmonium states 
$D_s^+ D_s^-$ pair in the hadronic medium.
The medium modifications of the $D_s$ mesons are
expected to be considerably enhanced at large densities 
and hence, the experimental consequences may be accessible 
in the upcoming CBM experiment, at the future facilities 
of FAIR, GSI \cite{CBM_Physics_Book}.

\end{section}  

\begin{section}*{Conflict of Interests} 
The authors declare that there is no conflict of interest regarding the publication of this paper. 
\end{section}

\begin{section}*{Acknowledgments} 
D.P. acknowledges financial support from University Grants Commission, India [Sr. No. 2121051124, Ref. No. 19-12/2010(i)EU-IV]. A.M. would like to thank Department of Science and Technology, Government of India (Project No. SR/S2/HEP-031/2010) for financial support. \\ \ 

\end{section}



\end{document}